\documentclass[letterpaper,onecolumn,12pt]{quantumarticle}
\pdfoutput=1
\usepackage[utf8]{inputenc}
\usepackage[english]{babel}
\usepackage[T1]{fontenc}
\usepackage{amsmath,amssymb}
\usepackage{caption}
\usepackage{subcaption}

\usepackage{tikz}
\usepackage{lipsum}
\usepackage{graphicx,multirow}
\usepackage{color,outlines}
\usepackage{hyperref}
\usepackage{physics}
\usepackage{outlines}

\def\be{\begin{equation}}
\def\ee{\end{equation}}
\def\bea{\begin{eqnarray}}
\def\eea{\end{eqnarray}}
\def\f{\frac}

\def\nn{\nonumber}

\def\ra{\rangle}
\def\la{\langle}

\begin{document}

\title{ Non-Hermitian description of sharp quantum resetting}

\author{Ranjan Modak}
\email{ranjan@iittp.ac.in} 
\author{S. Aravinda}
\email{aravinda@iittp.ac.in} 
\affiliation{Department of Physics, Indian Institute of Technology Tirupati, Tirupati, India~517619} 
\begin{abstract}
We study a non-interacting quantum particle, moving on a one-dimensional, which is subjected to repetitive  measurements. We investigate the consequence when such motion is interrupted and restarted from the same initial configuration, known as the quantum resetting problem. We show that such systems can be described by 
the time evolution under certain time-dependent non-Hermitian Hamiltonians. We construct two such Hamiltonians and compare the results with the exact dynamics. Using this effective non-Hermitian description we evaluate the timescale of the survival probability as well as the optimal resetting time for the system.  
\end{abstract}

\maketitle

\section{Introduction}
 Reaching the target at an optimal time is the most critical aspect in all branches of science, technology, and economics. The tools and methods to optimize the time to reach the target are well-studied in classical stochastic dynamics~\cite{redner2001guide,metzler2014first}. The significant discovery in the recent decade in classical stochastic dynamics is to restart (reset) the process to expedite the hitting time~\cite{evans2020stochastic,gupta2022stochastic}. In the classical randomized algorithmic studies, particularly Los Vegas algorithms, it has been shown that the resetting process optimizes the termination of the algorithm (completion). 

Quantizing the classical models has provided great insight and spurred progress in the technological aspects. The inherent randomness in quantum mechanical formalism offers an opportunity and a platform for a natural question to be set. Technologically, the emergence of the noisy intermediate-scale quantum (NISQ)-era computers and quantum thermodynamics necessitates all the tools to optimize the processes. Measurement disturbance is the  fundamental difference between classical and quantum dynamics which plays its role in studying the first detection process and any modifications. 

The natural question addressed in the classical stochastic resetting scenario is to optimize the  mean first passage time. Studies on `Time' in quantum theory have been debated from its inception~\cite{muga2007time}. The various notions like Mean First passage time, first arrival time distribution, first detected time, etc has been debated extensively~\cite{lumpkin1995extending,kumar1985quantum,schild2018time,sharma2012first,roncallo2022does,
muga2000arrival,allcock1969time,aharonov1998measurement,grot1996time,das2021questioning,giovannetti2015quantum}. Our work is concerned with the {\it first detected time} of the particles under the quantum evolution as explored in the Ref~\cite{dhar2015quantum,dhar2015detection,barkai.2017,dubey2021quantum,kessler2021first}. In a parallel study,   the random resetting of the quantum state to its initial state is been studied as a process to generate non-equilibrium steady states, and the dynamics can be modeled as an open quantum system dynamics~\cite{wald2021classical,rose2018spectral,mukherjee2018quantum,perfetto2021designing,perfetto2022thermodynamics,turkeshi2022entanglement,mogani2022emergent,das2022quantum,dattagupta2022stochastic,das2022quantuma}.

In previous studies, ~\cite{dhar2015detection,dhar2015quantum} it has been shown that the problem of a quantum particle moving on a lattice and subjected to repetitive measurements, can be effectively described by the evolution under non-Hermitian Hamiltonians. Non-Hermitian Hamiltonians have been a subject of interest for the last few decades in 
various branches of physics including open quantum systems, scattering theory, optics, and biological systems~\cite{Ozdemir_2019,shukla2022uncertainty,pal2022dna,HatanoNelson1996PRL,HatanoNelson1997PRB,car_93}, such systems have been experimentally realized as well~\cite{Yu_2020,dmi_16}. 
Very recently in Ref.~\cite{yin2022restart,eli2}, 
a similar problem of quantum detection has been studied within the quantum resetting setup. 
Essentially the goal was to eliminate the dark states in the quantum first-detection problem~\cite{eli3}.
It has been argued that with an optimal resetting setup, the quantum walker performs much better than the classical one, in the sense that the detection probability for the former is much higher in comparison to the classical case for a given time. 
Here, our goal is to see whether a non-Hermitian description can be found for the quantum resetting problems as well. We answer the question affirmatively and show that indeed one can construct time-dependent non-Hermitian Hamiltonians that describe the quantum measurement problem within a sharp resetting set-up efficiently.

\section{Quantum First Detection process\label{sec:qfd}}
The first detection of a quantum particle along a lattice is fundamentally different than its counterpart in the classical dynamics \cite{dhar2015quantum,dhar2015detection,barkai.2017,dubey2021quantum}. In classical mechanics, the state is unaffected by the measurement but in quantum theory the measurement disturbs the state and thereby alters the trajectory. In otherwords, the measurement is a non-unitary process. 
Consider that the dynamics of a quantum system in an one dimensional lattice of length $L$ is governed by the unitary operator $U= e^{-iHt}$, where $H$ is the Hermitian operator called Hamiltonian  of the system.  We are interested in measuring the statistics of particle detection at an arbitrary site $D=  \ketbra{s}$.  The measurement $\mathcal{M}$ is defined as a set $\{ D, I-D \}$, where $I$ is an identity operator. For simplicity, let's denote $B= I- D$.  If the state of the system is $\ket{\Psi}$, then the probability of detecting the particle at  $s$ site is $p = |\bra{s}\ket{\Psi}|^2 = \expval{D}{\Psi}$, and   the probability of non-detection (survival probability) is $P = 1-p = \expval{B}{\Psi}$. 

Let the initial state at time $t=0$ be $\ket{\Psi_0}$, and the state evolves under the unitary operator $U_\tau$ at discrete time steps $\tau$. The measurement is carried at every time instant $n\tau$, where $n = 1,2,\cdots N$.  For quantum detection,  $\tau$ needs to be non-zero.  Otherwise, $\tau\to 0$ leads
to freezing of the dynamics, and hence nothing will be detected.  This is known as the quantum Zeno effect~\cite{misra1977zeno}.  If the measurement happens at time $t = n \tau$, let the unnormalized state be $\ket{\Psi_n^{\pm\epsilon}} $  at $t = n \tau \pm \epsilon$.  $\pm \epsilon$ indicates the small time just after and before the measurement, respectively. Corresponding states at $t = n\tau$ is  
\begin{equation}
\ket{\Psi_n^{+\epsilon }} = \tilde{U}^n \ket{\Psi_0}, \quad \ket{\Psi_n^{-\epsilon }} = U_\tau \Tilde{U}^{n-1} \ket{\Psi_0}. 
\label{eq:pmevolv}
\end{equation}
Note that $\tilde{U} = B U_\tau$ and it is not unitary. 

The survival probability at time $t = n\tau$ is 
\begin{equation}
    P_n = \braket{\Psi_n^{+\epsilon}} = \expval{\tilde{U}^\dagger \tilde{U}}{\Psi_0}. 
\end{equation}
Hence, the probability of first detection  at $n$th measurement  can thus be estimated as:
\begin{equation}
p_n = P_{n-1} - P_n  = \expval{D}{\Psi_n^{-\epsilon }}. 
\label{eq:fdtp}
\end{equation} 
The amplitude and probability of the quantum first detection  is $\Psi_n = \braket{s}{\Psi_n^{-\epsilon}}$ and $p_n = |\Psi_n|^2$, respectively. The evolution of $\Psi_n$ is governed by the following {\it quantum renewal equation} 
~\cite{barkai.2017},
\begin{equation}\label{eq:renewal}
\Psi_n =\langle s|U(n\tau)|\Psi_0\rangle - \sum_{m=1}^{n-1} 		\langle s|U[(n-m)\tau]|s\rangle 
            \Psi_{m}.
\end{equation}
It is clear from Eqn.~(\ref{eq:renewal}) that the first-hitting amplitude is equal to the measurement-free transition amplitude 
$\langle s|U(n\tau)|\Psi_0\rangle $ 
subtracting the measurement-free return amplitude 
$\langle s|\hat{U}[(n-m)\tau]|s\rangle$
propagating from the prior first-hitting amplitude $\Psi_m$ ($m<n$). Finally, the integrated detection probability up to time $n\tau$ is given by $P_{{\rm det} }(n) = \sum_{n^\prime=1} ^n p_{n^\prime}$.

\section{Model and Protocols \label{sec:model}}
We study a system of fermions in a one-dimensional lattice of size $L$ ($L$ is chosen to be an even number) described by  the following tight-binding Hamiltonian:
\begin{eqnarray}
{H}&=&\sum_{j=1}^{L-1}({c}^{\dag}_j{c}_{j+1}+\text{H.c.}) \nonumber \\
\label{nonint_model}
\end{eqnarray}
where ${c}^{\dag}_j$  (${c}_j$) is the Fermionic creation (annihilation) operator at site $j$, ${n}_j ={c}^{\dag}_j{c}_{j}$ is the number operator.  We choose our initial state to be $|\Psi_0\ra={c}^{\dag}_{L/2}|0\ra$ (electron is located at the $\frac{L}{2}$-th site in the lattice), let it evolve under the Hamiltonian $H$.   
 We set a detector at $\ket{s}$ site, the unitary evolution $U_\tau=e^{-iH\tau}$ is repeatedly interrupted by projective measurements $D= \ketbra{s} $ at time interval $\tau$, $2\tau$, $3\tau$, $\cdots$. 
 
We follow the sharp restart strategy, and whenever $t=t_r = r\tau$, we set the system to its initial state $\ket{\Psi_0}$. If the particle is detected before the restart time, the process stops and calculates its time of first detection, else the particle evolves till retsart time and reset to its initial state.  If $t_f$ is the first detection time,  the mean first detection time  (MFDT) under restart can be expressed as $\expval{t_f}_r = \tau \expval{n_f}$ with $n_f = rR+\tilde{n}$. $n_f$ is the number of measurements carried until first detecting the particle with  $R \ge 0$ is the number of restarts, and  $1\le \tilde{n}\le r$. By definition, $R=[(n-1)/r]$ with $[...]$ means the integer part, and $\tilde{n}=n-r R=1+\text{mod}(n-1,r)$. Now, the quantum first detection probability with restart can be written as: 
\begin{equation}
    p_n^{(r)} = ( 1-P_\text{det}(r) )^R p_{\tilde{n}}.
\end{equation}
Consequently, the MFDT reads as~\cite{pal.2021},  
\begin{eqnarray}
    \langle t_f\rangle_r = \sum_{n=1}^\infty n\tau F_n^{(r)} 
    =  r\tau \frac{\left[ 1- P_\text{det}(r) \right] } {P_\text{det}(r)}
+\sum_{\tilde{n}=1}^r {(\tilde{n}\tau) \frac{p_{\tilde{n}}}  {P_{\rm det} (r)} }. \nonumber \\
\label{eq5}
\end{eqnarray}
We consider a sharp restart technique in the present study because of two reasons: (a) it is found to be most efficient in reducing the completion time of a search process in both classical and quantum mechanical setup~\cite{pal.2021,yin2022restart}, and (b) it is easy to  implement in a dynamics with discrete time steps. 

\section{Effective non-Hermitian model}
Unitary evolution followed by the measurement can be modeled as time evolution under an effective non-Hermitian Hamiltonian. 
They are not unique. In the following subsections, we discuss how to
construct such non-Hermitian Hamiltonian in two different ways. 

\subsection{Model I}
Given the detector is placed at site $s$, with detection operator $D=|s\ra \la s|$ and the complementary operator $B=I-D$, and  the Hamiltonian in Eq.~(\ref{nonint_model}) can be rewritten as,  
\begin{align}
H&=H_S+V~, \label{ham2} \\
{\rm where}~~H_S&=\sum_{j\neq {s,s-1}}({c}^{\dag}_j{c}_{j+1}+\text{H.c.}),~~~ {\rm and} ~~~
 ~~V= ({c}^{\dag}_{s-1}{c}_{s}+{c}^{\dag}_{s}{c}_{s+1}+\text{H.c.}). \nn
\end{align}
Expanding the effective evolution operator $\bar{U} = B e^{-iH t} B$ to second  order in $\tau$ one gets~\cite{dhar2015detection},
\begin{align}
\bar{U} &= B~\left[~I-iH \tau -\f{\tau^2}{2} H^2+\ldots~\right]~B \nn \\
&=I-iH_S \tau -\f{\tau^2}{2} H_S^2 
-\f{\tau^2}{2} \sum_{l,m}  
V_{l,s}V_{s,m} {c}^{\dag}_{l}{c}_{m} + \ldots \nn \\
&=e^{-i H^{0}_{eff} \tau}+{\mathcal{O}}(\tau^3), \nn \\
{\rm where}~~H^{0}_{eff}&=H_S+V_{eff}, \label{effH} 
{\rm and}~~V_{eff}=-\f{i \tau}{2} \sum_{l,m} V_{l,\alpha}
V_{\alpha,m} {c}^{\dag}_{l}{c}_{m} ~. \nn
\end{align}
Given the only non-zero elements of the matrix, $V$ are 
$V_{s,s-1}=V_{s-1,s}=V_{s,s+1}=V_{s+1,s}=1$, the effective 
non-Hermitian is given by, 
\begin{eqnarray}
     H^{0}_{eff}=\sum_{j\neq {s,s-1}}({c}^{\dag}_j{c}_{j+1}+\text{H.c.})&-\frac{i\tau}{2}({c}^{\dag}_{s-1}{c}_{s+1} 
     +\text{H.c.})-&\frac{i\tau}{2}(n_{s-1}+n_{s+1})
\label{heffI}
\end{eqnarray}
We will refer to this effective non-Hermitian Hamiltonian as model I in the rest of the manuscript.

\subsection{Model II}
The second model, which we call as Model II, in which a non-Hermitian Hamiltonian with a {\emph{large}} imaginary potential~\cite{krapivsky2014survival,dhar2015detection} is used. We follow the formalism derived in Ref.~\cite{krapivsky2014survival}, according to which the effective Hamiltonian contains a non-Hermitian on-site term 
$-i \Gamma$ at the site where the detector is placed ($\Gamma$ being a dimensionless and positive 
real number) which leads to a non-unitary evolution. Assuming that $\Gamma 
\gg 1$  and $\tau \ll 1$, one can  
develop a second-order perturbation theory in these quantities and show 
that the effective non-Hermitian can mimic 
the system if $\tau \Gamma = 2$.  The effective Hamiltonian is given by, 
\begin{align}
H^{0}_{eff}= H+ \Gamma H' ~~~{\rm{where}}~~ H'=- i{c}^{\dag}_s{c}_{s}, 
\end{align}
where $H$ is the Hamiltonian defined in Eq.~(\ref{nonint_model}).
The effective non-Hermitian Hamiltonian is given by, 
\begin{align}
H^{0}_{eff}= \sum_j({c}^{\dag}_j{c}_{j+1}+\text{H.c})-\frac{2i}{\tau}{c}^{\dag}_s{c}_{s}.
\label{heffII} 
\end{align}
\begin{figure*}
\subfloat[\label{fig1a}]{%
  \includegraphics[scale=0.3]{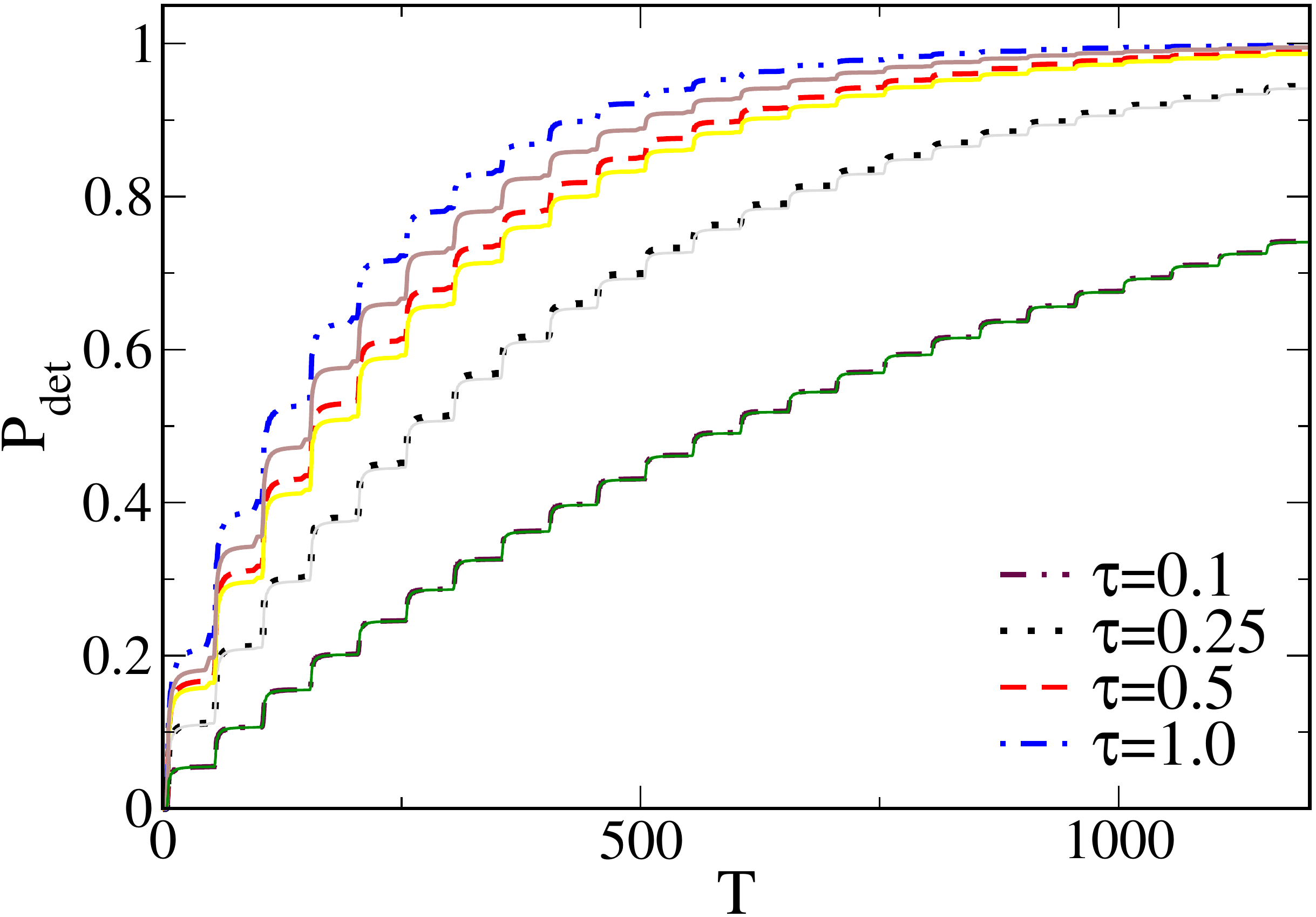}%
}\hfill
\subfloat[\label{fig1b}]{%
  \includegraphics[scale=0.3]{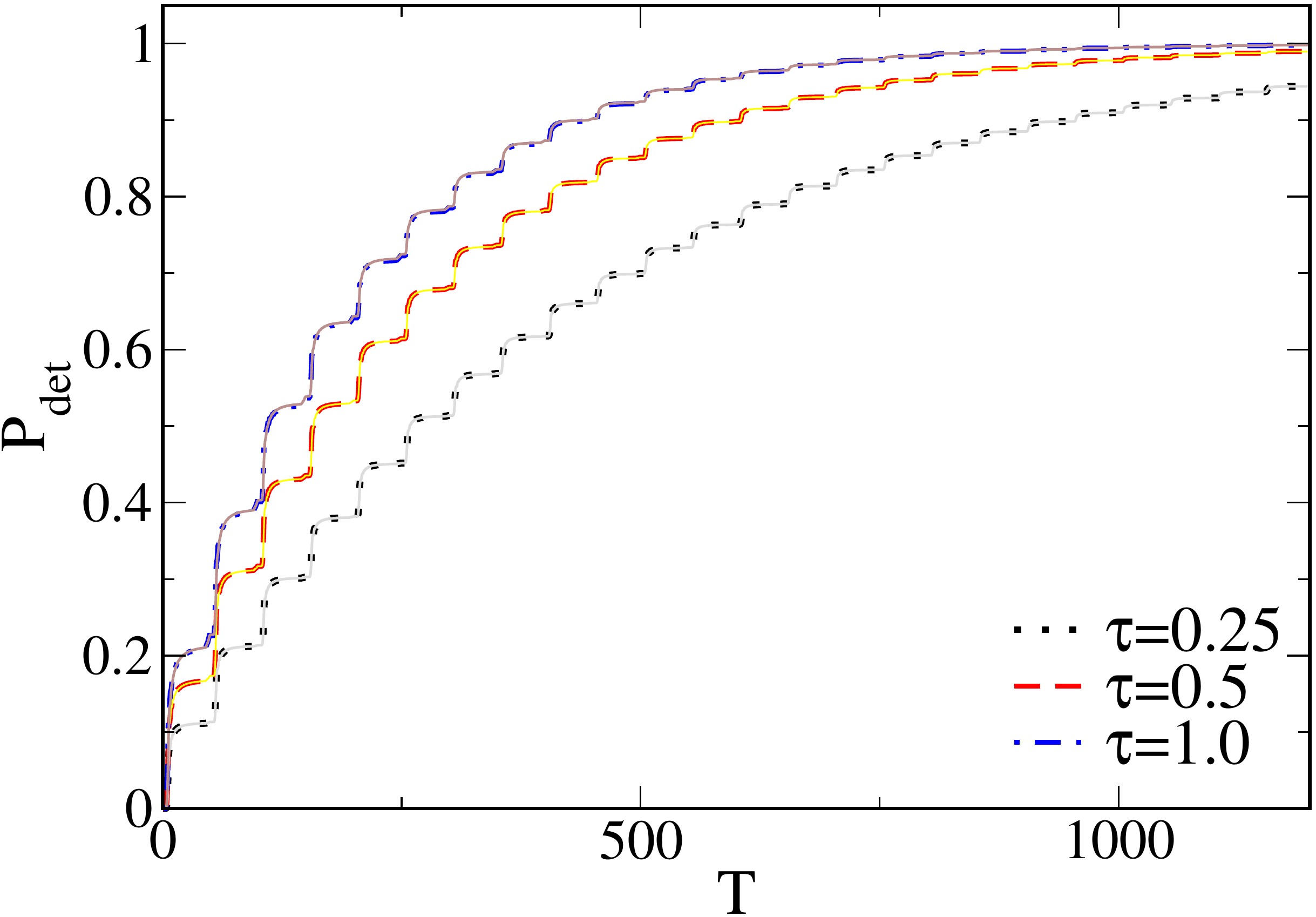}%
} 
\caption{(a) and (b) show the variation of $P_{det}$ with time $T$  for different values of $\tau$. Broken lines correspond to exact dynamics. Solid lines in (a) 
and (b) correspond to non-Hermitian dynamics for the model I and model II respectively.}

\label{fig1}
\end{figure*}

\section{Effective non-Unitary evolution for resetting}

In this section, our goal is to derive the effective Hamiltonian for  quantum resetting. We assume that resetting time is $t_r$, given that without resetting the system can be described by a non-unitary evolution where underlying effective Hamiltonian is non-Hermitian (\eqref{heffI} and \eqref{heffII}), the survival probability $P$ at time $t_r$ is given by (assuming the initial state $\Psi_0\ra$),
\begin{equation}
    P(t_r)=\la\Psi_0| e^{iH^{\dagger}_{eff}t_r}e^{-iH_{eff}t_r}|\Psi_0\ra. 
\end{equation}
The survival probability after a time $t_rR+t$ ($R$ represents the number of time resetting has already taken place), where $t<t_r$, is given by, 
\begin{equation}
    P(Rt_r+t)=\la\Psi_0| e^{iH^{\dagger}_{eff}t}e^{-iH_{eff}t}|\Psi_0\ra [P(t_r)]^R. 
\end{equation}

\begin{figure*}
\subfloat[\label{fig2a}]{%
  \includegraphics[scale=0.3]{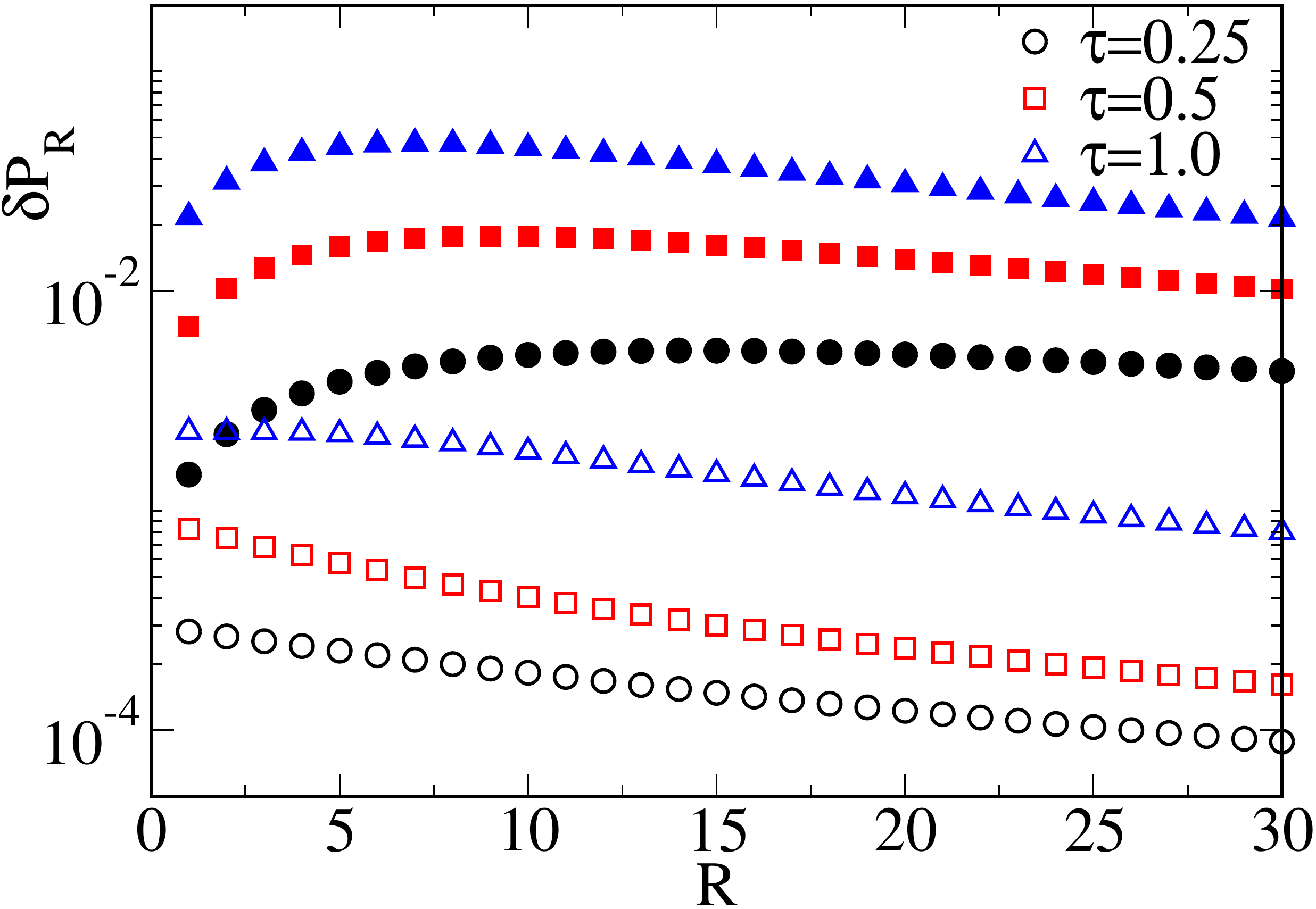}%
}\hfill
\subfloat[\label{fig2b}]{%
  \includegraphics[scale=0.3]{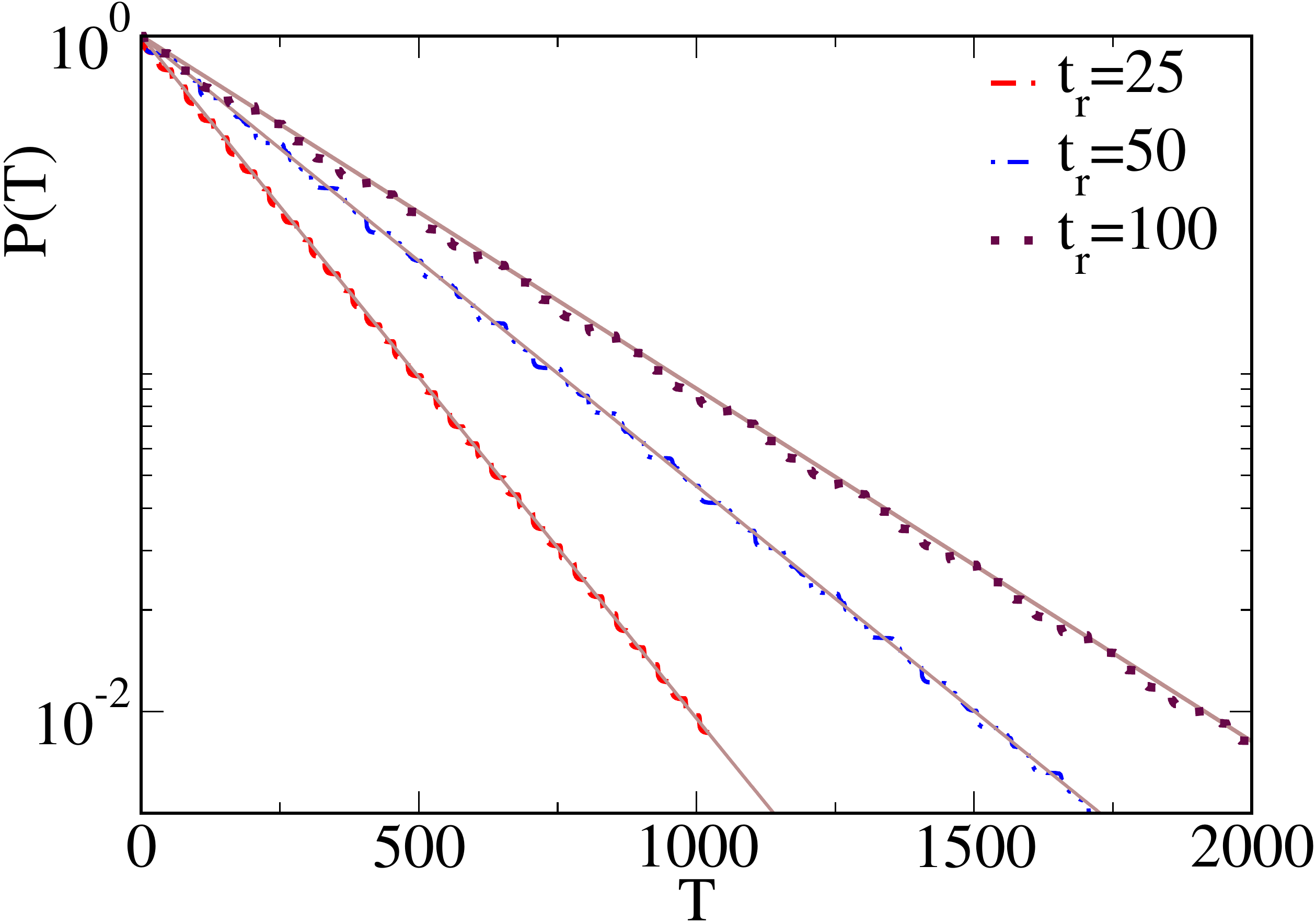}%
} 
\caption{(a) shows the variation of $\delta P_R$ (see Eqn.~\eqref{dpr}) with $R$ (number of resetting) for 
different values of $\tau$. Open symbols (filled symbols) correspond 
to $\delta P_R$ where $H_{eff}$ is taken to be model II (model I). (b) shows the variation of survival probability $P(T)$ with time $T$ for different resetting time $t_r$ and $\tau=0.25$. Solid lines are analytical prediction $P(T)\sim e^{-T/T_s}$, where $T_s=t_r\alpha^{-1}$ is obtained from the model II.}
\label{fig2}
\end{figure*}

Let's first consider the $R=1$ case. The survival probability at time $t_r+t$ is 
$P(t_r+t)=\la\Psi_0| e^{iH^{\dagger}_{eff}t}e^{-iH_{eff}t}|\Psi_0\ra [P(t_r)]=\la\Psi_1| e^{iH^{\dagger}_{eff}t}e^{-iH_{eff}t}|\Psi_1\ra$,
 where $|\Psi_1\ra=\sqrt{P(t_r)}|\Psi_0\ra$. One can re-write this as, 
 \begin{equation}
    P(t_r+t)=\la\Psi_0| e^{iH^{1\dagger}_{eff}t}e^{-iH^{1}_{eff}t}|\Psi_0\ra, \nn  
\end{equation}
where
\begin{eqnarray}
    H^{1}_{eff}=H^{0}_{eff}+\frac{i}{2t}\ln |\la\Psi_0| e^{iH^{0\dagger}_{eff}t}e^{-iH^0_{eff}t}|\Psi_0\ra|\sum_{j}n_j. \nn\\
\end{eqnarray}
Similarly, it is straightforward to check that, for arbitrary $R$ and at a time $Rt_r+t$ the survival probability reads as,  
\begin{equation}
    P(Rt_r+t)=\la\Psi_0| e^{iH^{R\dagger}_{eff}t}e^{-iH^{R}_{eff}t}|\Psi_0\ra \nn,  
\end{equation}
where
\begin{eqnarray}
    H^{R}_{eff}=H^{0}_{eff}+\frac{iR}{2t}\ln |\la\Psi_0| e^{iH^{0\dagger}_{eff}t}e^{-iH^0_{eff}t}|\Psi_0\ra|\sum_{j}n_j. \nn\\ 
\end{eqnarray}
Assuming $\alpha=-\ln |\la\Psi_0| e^{iH^{0\dagger}_{eff}t}e^{-iH^0_{eff}t}|\Psi_0\ra|>0$,  the survival probability at time $Rt_r+t$ can be written as, 
\begin{equation}
    P(Rt_r+t)=\la\Psi_0| e^{iH^{0\dagger}_{eff}t-\frac{R\alpha}{2}\sum_jn_j}e^{-H^{0}_{eff}t-\frac{R\alpha}{2}\sum_jn_j}|\Psi_0\ra.\nn 
\end{equation}
Given $[H^{0}_{eff},\sum_jn_j]=[H^{0\dagger}_{eff} , \sum_jn_j]=0$, 
\begin{eqnarray}
      P(Rt_r+t)&=&\la\Psi_0| e^{iH^{0\dagger}_{eff}t}e^{-H^{0}_{eff}t} e^{-R\alpha\sum_jn_j}|\Psi_0\ra \nn\\
    &=&e^{-\alpha R}\la\Psi_0| e^{iH^{0\dagger}_{eff}t}e^{-H^{0}_{eff}t}|\Psi_0\ra. \nn\\
\end{eqnarray}
  This ensures that in the limit $R\to \infty$,  $P(Rt_r+t) \to 0$, which implies $P_{det}\to 1$.  It also proves that the survival probability goes to zero with increasing time $T$ as $P(T)\sim e^{-T/T_s}$, where 
 $T_s=t_r\alpha^{-1}$ can be identified as a survival timescale.   

Given that $T_s$ provides a time scale for survival,   the optimal resetting time $t^*_r$ is the one, for which $T_s$ is the minimum. It implies, 
\begin{eqnarray}
    t^*_r=\min_{t_r}~~\frac{1}{t_r}\ln |\la\Psi_0| e^{iH^{0\dagger}_{eff}t}e^{-iH^0_{eff}t}|\Psi_0\ra|.
    \label{tr_con}
\end{eqnarray}

\begin{figure*}
\subfloat[\label{fig3a}]{%
  \includegraphics[scale=0.3]{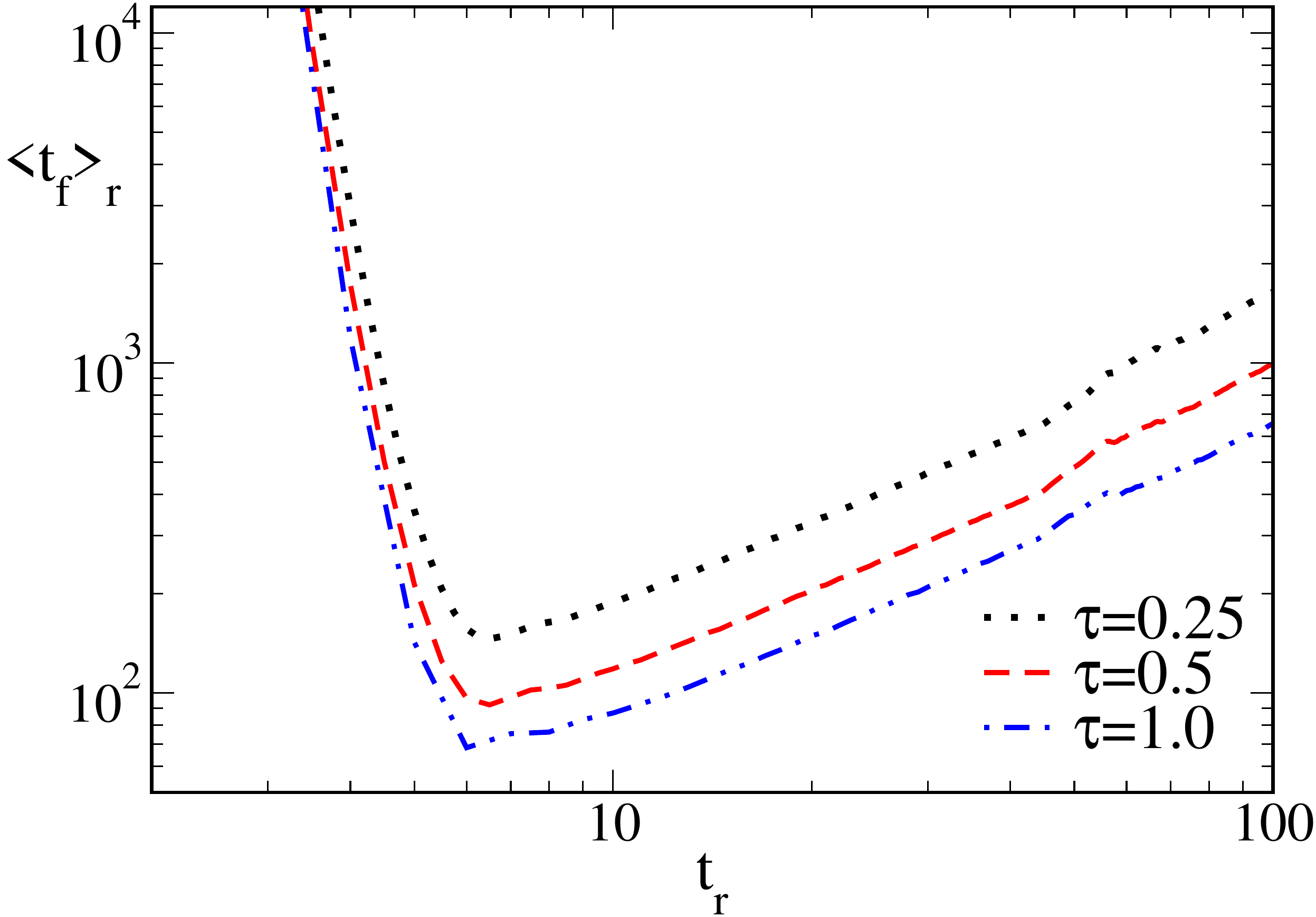}%
}\hfill
\subfloat[\label{fig3b}]{%
  \includegraphics[scale=0.3]{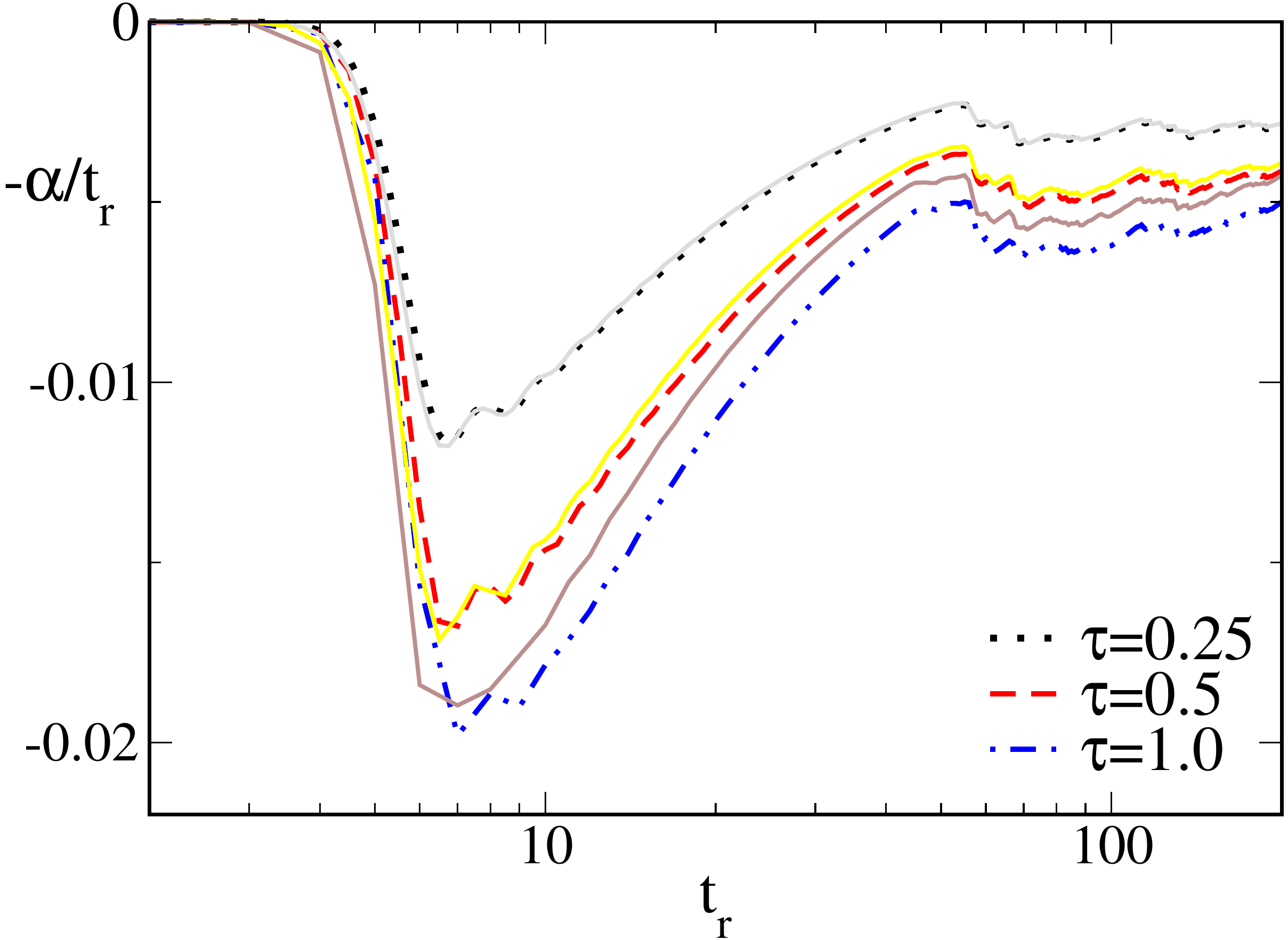}%
} 
\caption{(a) shows the variation of MFDT with $t_r$  for different values of $\tau$. (b) shows the variation 
$-\alpha/t_r$ with $t_r$ for different values of $\tau$, where $\alpha$ is obtained from the effective models. Broken lines correspond to model II and solid lines correspond to model I. }
\label{fig3}
\end{figure*}
 
 \section{Numerical Results}

   First, we compare our exact results with two effective  non-Hermitian Hamiltonians (model I and model II). For all our numerical calculations in this manuscript, we choose $L=500$, and the position of the detector is set to be at the site $s=L/2+10$. 
 Given it has been argued  that the non-Hermitian Hamiltonians constructed here, are effective descriptions of the exact dynamics only in the small $\tau$ limit, we compare our results in Fig.~\ref{fig1} for different values of $\tau$ and plot $P_{det}$ vs $T$. We find that indeed for small $\tau$, the non-Hermitian dynamics can mimic the exact dynamics brilliantly. However, as $\tau$
   increases the comparison becomes worse.  
   In order to quantify it, we define a quantity, 
\begin{eqnarray}
 \delta P_R(R)=\frac{\tau}{t_r}\sum_{(R-1)t_r < T\le Rt_r} |P_{det}^{exact}-P_{det}^{H_{eff}}|.
 \label{dpr}
\end{eqnarray}
Now depending on how small $\delta P_R(R)$ is, we can conclude our effective non-Hermitian description is that much accurate. Figure.~\ref{fig2a} clearly demonstrate that $\delta P_R$ decreases as we increase $\tau$. That is expected from the construction of $H_{eff}$. However, Fig.~\ref{fig2a} also suggests that model II works better compared to the model I, given for fixed $\tau$,  $\delta P_R$ for model II (data shown in open symbols) are much smaller than the model I data (data shown in filled symbols). Next, we also 
study how the survival probability behaves with time. 
Figure.~\ref{fig2b} shows the variation of the survival probability $P(T)$ with time for different values of resetting time $t_r$. It shows an exponential decay 
$P(T)\simeq e^{-T/T_s}$ (results in Fig.~\ref{fig2b} shown in the semi-log scale). Using the effective non-Hermitian description, we predicted in the previous section that the survival time-scale $T_s=t_r\alpha^{-1}$. Solid lines in Fig.~\ref{fig2b}
correspond to $P(T)\simeq e^{-\alpha T/t_r}$ (where $\alpha$ is obtained using model II), they match brilliantly with the exact dynamics. 

Finally, we study the mean first detection time (MFDT)
and estimate optimal resetting time. While studying 
exact dynamics, MFDT can be calculated using Eqn.~\eqref{eq5}. The optimal reset time $t^*_r$ corresponds to the resetting for which MFDT is minimum. The question is whether one can obtain $t^*_r$ from the effective non-Hermitian dynamics. In the previous section, we derived a condition to obtain $t^*_r$ from the non-Hermitian dynamics (see Eqn.~\eqref{tr_con}), Fig.~\ref{fig3} is a numerical validation of our finding. Figure.~\ref{fig3a} shows the variation MFDT 
with $t_r$ for different values of $\tau$. We find that the minimum occurs around $t^*_r \simeq 6.5$, which corresponds to the optimal resetting time. Figure.~\ref{fig3b} shows the variation of $-\alpha/t_r$ with $t_r$ for both effective models. Indeed, we find that the minimum corresponds once again around $t^*_r\simeq 6.5$. Interestingly, in Fig.~\ref{fig1a}, it was observed that for $\tau=1$, model I fails to describe the exact dynamics efficiently, however, it still predicts optimal resetting time quite efficiently even for $\tau=1$. 
\section{Conclusions} 
We show that the quantum detection within a resetting 
set-up can be modeled as time evolution under a time-dependent non-Hermitian Hamiltonian. We have demonstrated it for one-dimensional tight-binding Hamtiontian where the detector was placed at a single site. Such effective non-Hermitian Hamiltonian is not unique, we have constructed two such Hamiltonians and compared them with exact dynamics.  While by construction the effective description is much more accurate in small $\tau$ limits, surprisingly we find one of them (which we refer to as model II) works much better than the other (model I) even relatively large $\tau$. These non-Hermitian descriptions allow us 
to predict the survival time scale analytically and it is also straightforward to see that without resetting the probability of detection is less than $1$ but resetting ensures confirmed detection in the long time limit. Moreover, we also demonstrate how the optimal resetting time can be computed from the effective non-Hermitian dynamics. 

In the future, it may be interesting to look at many-body interacting systems and investigate the effect of quantum resetting in the context of repetitive  quantum measurements~\cite{modak2021finite}. Moreover, this effective description 
can be extremely useful to understand the role of entanglements in quantum resetting and  also to study the effect of measurements of observable other than
the position.

\acknowledgements
RM acknowledges the DST-Inspire fellowship by the Department of Science and Technology, Government of India, SERB start-up grant (SRG/2021/002152). SA acknowledges the start-up research grant from
SERB, Department of Science and Technology, Govt. of India (SRG/2022/000467). The authors Debasish Mondal and Somrita Ray for the discussion.  The authors also thank Diptiman Sen for his comments on the manuscript. 
\bibliographystyle{unsrt}
\bibliography{reset} 

\newpage
\appendix

\end{document}